\documentclass[pre,preprint,showpacs,showkeys,nofootinbib,preprintnumbers,amsmath,amssymb,floatfix]{revtex4-1}

\usepackage{graphicx}
\usepackage{amssymb}
\usepackage{color}
\usepackage{epsfig}

\newcommand{\br}{{\bf r}}
\newcommand{\bx}{{\bf x}}
\newcommand{\bk}{{\bf k}}
\newcommand{\bv}{{\bf v}}
\newcommand{\bu}{{\bf u}}
\newcommand{\bff}{{\bf f}}
\newcommand{\kt}{k_{\rm B}T}

\newcommand{\eq}[1]{Eq.~(\ref{#1})}

\begin{document}

\title{Hydrodynamic interactions induce anomalous diffusion under partial confinement}

\author{J.~Bleibel$^{1,5}$, A.~Dom\'\i nguez$^2$, F.~G\"unther$^3$,
  J.~Harting$^{3,4}$, M.~Oettel$^5$} 
\affiliation{$^1$Max-Planck-Institut f\"ur Intelligente Systeme,
  Heisenbergstr.~3, 70569 Stuttgart, Germany}\email{bleibel@is.mpg.de} 
\affiliation{$^2$F\'\i sica Te\'orica, Universidad de Sevilla, Apdo.~1065,
  41080 Sevilla, Spain}\email{dominguez@us.es}
\affiliation{$^3$Fakulteit Technische Natuurkunde, Technische Universiteit
  Eindhoven, Den Dolech 2, 5600MB Eindhoven, The Netherlands }
\affiliation{$^4$Institut f\"ur Computerphysik, Universit\"at
  Stuttgart, Allmandring 3, 70569 Stuttgart, Germany}
\affiliation{$^5$Institut f\"ur Angewandte Physik, Universit\"at T\"ubingen,
  Auf der Morgenstelle 10, 72076 T\"ubingen}

\date{\today}

\begin{abstract}

   {Under partial confinement, the {motion} of colloidal particles
     {is restricted} to a plane  {but their dynamics is} influenced by hydrodynamic
    interactions mediated by the unconfined, three--dimensional flow
    of the embedding fluid. We demonstrate that this dimensionality
    mismatch  
    induces a characteristic divergence in
    the two--dimensional collective diffusion coefficient of the
    colloidal monolayer. This result,  {independent from the specific interparticle forces in the monolayer,} 
   is solely due to the kinematical
    constraint on the colloidal particles,  {and it} is different from the known
    divergence of transport coefficients in purely  {two--dimensional} fluids.
  }

\end{abstract}



\maketitle

{The diffusive behavior of macroobjects in solution (colloids, micelles,
polymers ...)
{is of major importance}
for {addressing fundamental aspects of Statistical Physics and for applications}.
Diffusion governs the transport of particles in heterogeneous environments
often encountered in soft matter and biological systems, therefore
normal and anomalous diffusion has been studied by the {corresponding}
communities from different viewpoints and {with diverse} methods
\cite{Lippincott:2001,Metzler:2000}.  {A fundamental
  characterization of the diffusive dynamics is provided by the
  diffusion coefficients. For the simplest case of a dispersion of
  colloidal particles, the Fourier components $\delta\varrho_\bk(t)$
  of the particle number density field evolve according to
  \begin{equation}
    \label{eq:diff}
    \frac{\partial \delta\varrho_\bk}{\partial t} = - D(k) k^2 \delta\varrho_\bk 
  \end{equation}
  in the long--time, large--scale regime. This serves to define the
  wave number dependent \emph{coefficient of collective diffusion},
  $D(k)$, and the associated diffusion constant, $D(k\to 0)$.}
It is known that $D(k)$}
is influenced by (i) {static effects}, e.g., the effective
interactions between the macroobjects, and (ii) {genuinely dynamical
  effects}, in particular the hydrodynamic interactions (HI) {mediated
  by} the solvent, usually resulting from the overdamped regime {of
  motion} (Stokes flow).

The dynamics 
of colloidal solutions in confinement  {or near physical boundaries} has also been of long--standing
interest. 
 {The effect of HI has}
been investigated  {theoretically} on systems confined between
walls~\cite{Pesche:2000, Swan:2011}, close to a free
interface~\cite{RZMM99,Cichocki:2004}, or on solutions with a matrix
of fixed obstacles \cite{Hoef13}.
Trapped objects in (fluid) membranes also show peculiar diffusion
behavior~\cite{Saffman:1975,Ramachandran:2010}. In this letter, we investigate
the generic case 
of {\em partial confinement}:  {one part of the system, namely the
  colloidal particles are restricted to move in a two--dimensional (2D)
  plane}, whereas  {the other part, namely the} solvent below and
above the plane evolves in an (essentially unbound) three--dimensional
(3D) domain.
An obvious realization of such
a partial confinement setup  {is a colloidal monolayer}
at fluid interfaces 
-- here the particles are irreversibly trapped under partial wetting
conditions and their movement is thus restricted to the surface
defined by the sharp fluid interface.
{Here we study this problem {both} theoretically with a simplified
  model {and numerically} with 
  simulations of a more complete model of the monolayer and the
  embedding fluid. The main result 
  is that the HI induce \textit{anomalously fast diffusion}, showing
  up in the form of {a divergence $D(k\to 0) \sim 1/k$.}
}

 {\em Theory.--}  {We present first a theoretical model for the
  long--time, collective diffusion of a planar monolayer.
}  The number density $\varrho(\br, t)$
of particles at the plane (located at $z=0$), and the particle
velocity field $\bv(\br,t)$ are related by the continuity equation,
\begin{equation}
  \label{eq:cont}
  \frac{\partial \varrho}{\partial t} = - \nabla\cdot(\varrho \bv) ,
  \qquad
  \br=(x,y).
\end{equation}
The particles are acted upon by a force (external or due to direct
particle--particle interactions) and simultaneously transported by the
ambient flow in the surrounding 3D fluids. Thus, in the overdamped
limit we approximate
\begin{equation}
  \label{eq:v}
  \bv = \Gamma \bff + \bu ,
\end{equation}
where $\Gamma$ is the effective mobility of a particle  {at the
  interface}, $\bff(\br,t)$ is the average force per particle and
$\bu(\br,t)$ is the 3D ambient flow field evaluated at the plane
$z=0$. The force field $\bff(\br)$ is assumed to be expressible as a
functional of the density field: this includes many cases of physical
relevance (external forces, local thermal equilibrium), as will be
discussed below.
The ambient flow $\bu(\br)$, being induced by the motion of the
particles, is responsible for the HI
between the
particles and must be determined self--consistently as a function of
the force field $\bff(\br)$. We introduce the simplifying assumptions
that the dynamical viscosity $\eta$ has the same value for the fluid
above and below the interface and that the ambient flow is 3D
incompressible, laminar and smooth at the planar interface. The
deformation of the latter is assumed to be negligible. (Effectively,
we can dismiss any difference between the upper and the lower fluid
concerning the ambient flow for our purposes).
Under these conditions, the stationary ambient flow profile is
 {modeled as}
\begin{eqnarray}
  \label{eq:ambient}
  \bu(\br) &=& \frac{1}{8\pi\eta} \int d^2\br' \; \varrho(\br')\bff(\br') \cdot
  \mathcal{G}(\br-\br') , \nonumber \\ 
  & &\mathcal{G}_{\alpha\beta}(\br) = \frac{1}{|\br|} \left[ 
    \delta_{\alpha\beta} + 
    \frac{r_\alpha r_\beta}{|\br|^2} \right] .
\end{eqnarray}
Notice that $\mathcal{G}$ is the Oseen tensor \textit{for the 3D
  flow}, even though only its evaluation at points of the planar
interface ($z=z'=0$) is required. 
One can view this model as a mean--field--like approximation to the
effect of HI 
given that we make explicit only the
 {far--field} contribution that does not require a short--distance
cutoff but is, on the contrary, divergent in the infinite--size limit.
 {Actually, if the force $\bff (\br)= \nabla
  [\delta\mathcal{F}/\delta\varrho(\br)]$ can be derived from a free
  energy functional $\mathcal{F}[\varrho]$,
  Eqs.~(\ref{eq:cont})--(\ref{eq:ambient}) become a simple version of
  the dynamic density functional theory extended to include HI
  \cite{Rex:2009}.}
 {The approximations underlying
  Eqs.~(\ref{eq:cont})--(\ref{eq:ambient}) usually imply a restriction
  to the ``hydrodynamic'' regime (large scales, long times).}

In the absence of external force fields, the homogeneous, stationary
state, $\varrho(\br,t) = \varrho_\mathrm{hom}$, $\bff(\br, t) = 0$,
$\bu(\br, t) = 0$, is a possible solution of
Eqs.~(\ref{eq:cont})--(\ref{eq:ambient}). By linearizing them about the
homogeneous state, one obtains an equation for the evolution of the
perturbation $\delta\varrho(\br,t) = \varrho(\br, t) -
\varrho_\mathrm{hom}$,
\begin{equation}
  \label{eq:delta}
  \frac{\partial \delta\varrho}{\partial t} \approx
  - \Gamma\varrho_\mathrm{hom} \nabla\cdot\bff - 
  \varrho_\mathrm{hom} \nabla\cdot\bu .
\end{equation}
In this linear approximation, $\bff$ can be approximated as a linear
functional of $\nabla \delta\varrho$ in general, so that the first
term describes the decay (or growth, in cases of instability) of
density fluctuations driven by the force field. The second term
accounts for the effect of HI 
and the key point is to note that $\nabla\cdot\bu\neq 0$ in the 2D
layer, that is, the ambient flow at the plane $z=0$ induces
compression and dilution of the colloidal fluid, although the full 3D
ambient flow is not compressible. This is at variance with the
phenomenology when  {$\nabla\cdot\bu=0$} (absence of confinement
 {or full confinement to 2D}), under which conditions the effect of
the HI
appears  {only} as a nonlinear coupling (advection). 
By introducing the Fourier transform of the fields, \eq{eq:delta} can
be cast into the form of \eq{eq:diff} with {$D(k) = [ 1 + g(k) ]
  D_0 (k)$}.
Here $D_0(k)$, defined from
$\bff_\bk \approx [D_0(k) /\Gamma\varrho_\mathrm{hom}] (-i \bk
\delta\varrho_\bk)$, is the 
diffusion coefficient
in the absence of hydrodynamic couplings,  {the latter being
  accounted for by the function}
\begin{equation}
  \label{eq:g}
  g(k) = \frac{\varrho_\mathrm{hom}}{8\pi\eta\Gamma} 
  \sum_{\alpha,\beta=1}^2 \frac{k_\alpha k_\beta}{k^2}
  \int d^2\bx\; \mathrm{e}^{-i\bk\cdot\bx} 
  \mathcal{G}_{\alpha\beta} (\bx) 
  = \frac{1}{k L_\mathrm{hydro}} 
\end{equation}
in terms of a characteristic length $L_\mathrm{hydro} :=
4\eta\Gamma/\varrho_\mathrm{hom}$.
The behavior $g(k) \propto 1/k$ leads to the central result of our analysis: 
Since $g(k)>0$, the (linearized) evolution of a given mode
$\delta\varrho_\bk$ is accelerated, but its stability character,
determined by the sign of $D_0(k)$, is unchanged. The acceleration is
maximal for the largest spatial scales; in particular, since one
usually has  {$D_0(k\to 0) = \mathrm{finite}$}, 
our result leads to the conclusion that the collective diffusion is
anomalous,  {$D(k\to 0) = \infty$};
in real systems, this
divergence will be regularized by finite--size effects. The pole in
\eq{eq:g} is a consequence of computing the 2D Fourier transform of
the 3D Oseen tensor and can be traced back directly to the
 {kinematical} constraint imposed by partial confinement.
 {Contributions beyond the Oseen approximation
   {(see \eq{eq:ambient})}
   are expected to be subdominant in \eq{eq:g} as $k\to 0$; their
   effect would show up as finite corrections to the value of the
   mobility $\Gamma$.
}

The enhancement of diffusion is illustrated by means of two physically
relevant models lying at opposite extremes, namely,  {an ideal gas
  (particles do not exert a direct force on each other) and a gas of
  capillary charges (particles experience the extremely long--ranged
  capillary attraction, 
  reducing to 2D self--gravity in a limiting
  case~{\cite{Bleibel:L2011}})}. In the first 
case, $\bff = - \kt \nabla\ln\varrho$ and $D_0(k) = \Gamma \kt$.
The Green function $G(r,t)$ of
\eq{eq:delta}, obtained directly from the solution to \eq{eq:diff}, is
\begin{equation}
  \label{eq:green}
  G(r,t) = \frac{t_\mathrm{hydro}}{2\pi D_0 t^2} 
  \left[ 1 + 
    \left(\frac{r}{L_\mathrm{hydro}}\right)^2
    \left(\frac{t_\mathrm{hydro}}{t}\right)^2 
  \right]^{-3/2} ,
\end{equation}
 {for long times, $t\gg t_\mathrm{hydro} := L_\mathrm{hydro}^2/D_0$.}
In comparison with the diffusion without
HI, 
the density at the center ($r=0$) is reduced by a factor
$2t_\mathrm{hydro}/t$ and the decay at large distances is algebraic
instead of  {Gaussian}.

\begin{figure}[th]
  \centering \psfig{file=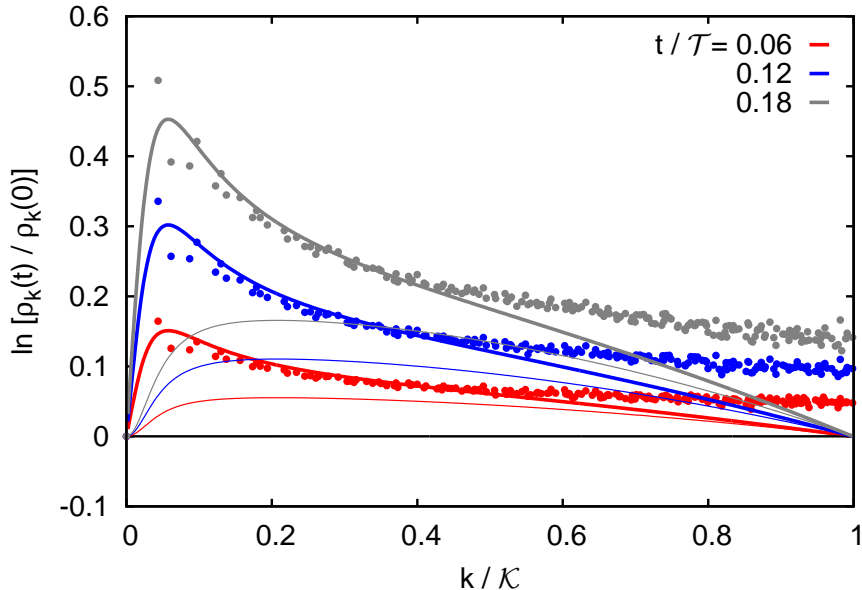,height=.7\linewidth,angle=270}
  \caption{Growth of 
    $\varrho_\mathbf{k}(t)$ 
    from tSD simulations (symbols; averaged over 2000 runs) and
     {from
      linear theory,
       {Eqs.~(\ref{eq:diff}, \ref{eq:Dcap})}
      (thick lines; thin lines in the absence of HI)}.
    Parameters in the tSD simulations: 3844  {hard spheres} (radius
    $10$ $\mu$m) in a box of size $7160\times 7160$ $\mu$m$^2$
    (periodic boundary conditions) with $\lambda=1100$ $\mu$m,
    $1/\mathcal{K}=49$ $\mu$m, $\mathcal{T}=52863$s.  
  }
  \label{fig:g_k}
\end{figure}

In the second case,  {the trapped particles deform the fluid
  interface and an effective mutual interaction of capillary origin
  arises (see, e.g., \cite{Kralchevsky:2000}). 
  In the simplest model (two--body
  force between capillary monopoles),}
the pairwise, attractive interaction potential is proportional to the
Bessel function $K_0(r/\lambda)$, dependent on the capillary length
$\lambda$. When $\lambda\to\infty$,
 {this reduces to} the Newtonian gravitational potential in 2D. For
realistic configurations ($\lambda\sim\mathrm{mm}$, typical colloidal
particle sizes $\sim\mu\mathrm{m}$), one can use the mean--field
approximation to compute the force $\bff$ and derive
\begin{equation}
  \label{eq:Dcap}
  D_0(k) = \frac{1}{\mathcal{T}} 
  \left( \frac{1}{\mathcal{K}^2} - \frac{1}{k^2 + 1/\lambda^2} \right) .
\end{equation}
 {Here, $\mathcal{T}$ 
  is a characteristic time scale
  and $1/\mathcal{K}$ 
  is a characteristic length scale; both depend on
  $\varrho_\mathrm{hom}$ and on temperature through properties of the
  fluid interface and the monolayer}
(see Ref.~\cite{Bleibel:2011} 
 for further details). According to
\eq{eq:Dcap}, the homogeneous state is unstable against clustering
 {($D(k)<0$ for some $k$)} below a given temperature,
and the dynamical evolution of the instability is dominated by the
large--scale modes,  {$k\ll \mathcal{K}$, for realistic values of
  the parameters}. As a consequence, the hydrodynamics--induced
acceleration can affect significantly the evolution  {(see thin and
  thick lines in Fig.~\ref{fig:g_k})}.

{\em Results from simulations.--} The preceding theoretical analysis
 {has been tested and extended beyond the linear regime by means of}
simulations of a colloidal monolayer. 
As our workhorse simulation method, we choose quasi-2D Brownian
dynamics (BD) simulations~\cite{Bleibel:2011}.  {We include HI,
  truncated at the two--body level, through}
the Rotne--Prager approximation, leading to a truncated Stokesian
dynamics  {(tSD) \cite{Brady:1988} that already incorporates
the Oseen tensor and thus
the physics discussed}
in the previous paragraphs. 
The 
tSD simulations  {are} validated 
{using a combined 3D multicomponent Lattice Boltzmann and Molecular Dynamics
  algorithm (LB)~\cite{Jansen:2011} 
  which includes HI}
at the many--body level. {{Using} these simulation techniques, we illustrate the effect of HI on two aspects of the
dynamics in the capillary collapse scenario and, as a third example, on the diffusional
behavior in a 2D ideal gas of colloidal particles:}  

$(i)$--HI--enhanced diffusion coefficient  {in capillary collapse:
  we check the linear prediction $\varrho_\bk(t) \propto \exp{(-D(k)
    k^2 t)}$ from \eq{eq:diff} for capillary
  monopoles} 
using tSD simulations. 
{As Fig.~\ref{fig:g_k} illustrates}, simulation results are described
very well by theory and clearly show the enhancement due to the $1/k$
divergence of the diffusion coefficient compared with the
corresponding results without HI. {(The deviations at large values
  of $k/\mathcal{K}$ {are due to nonlinear effects, and to
    corrections to mean--field from the short--range repulsion.})} 

\begin{figure}[th]
  \centering \epsfig{file=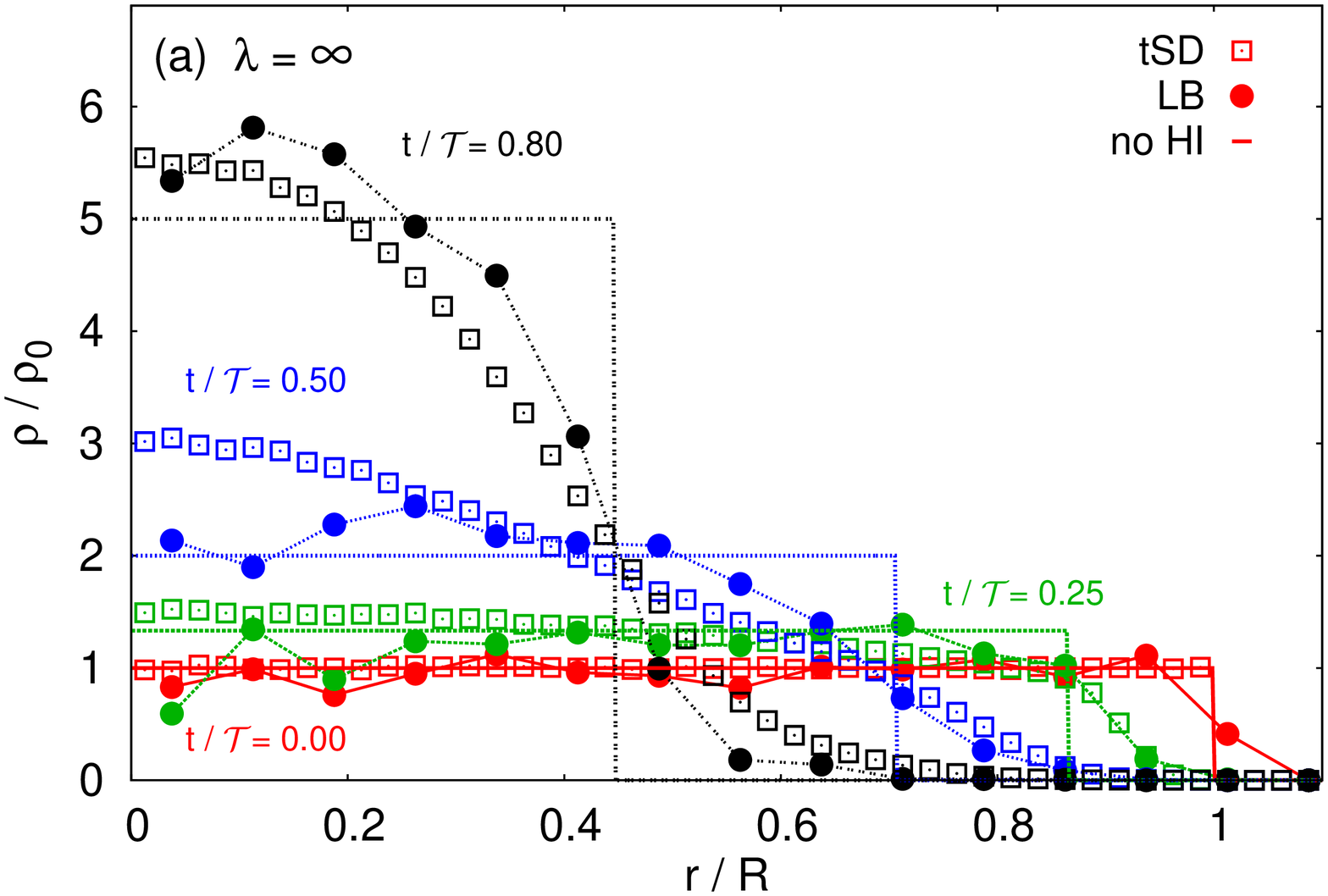,width=.7\linewidth}
  \centering \epsfig{file=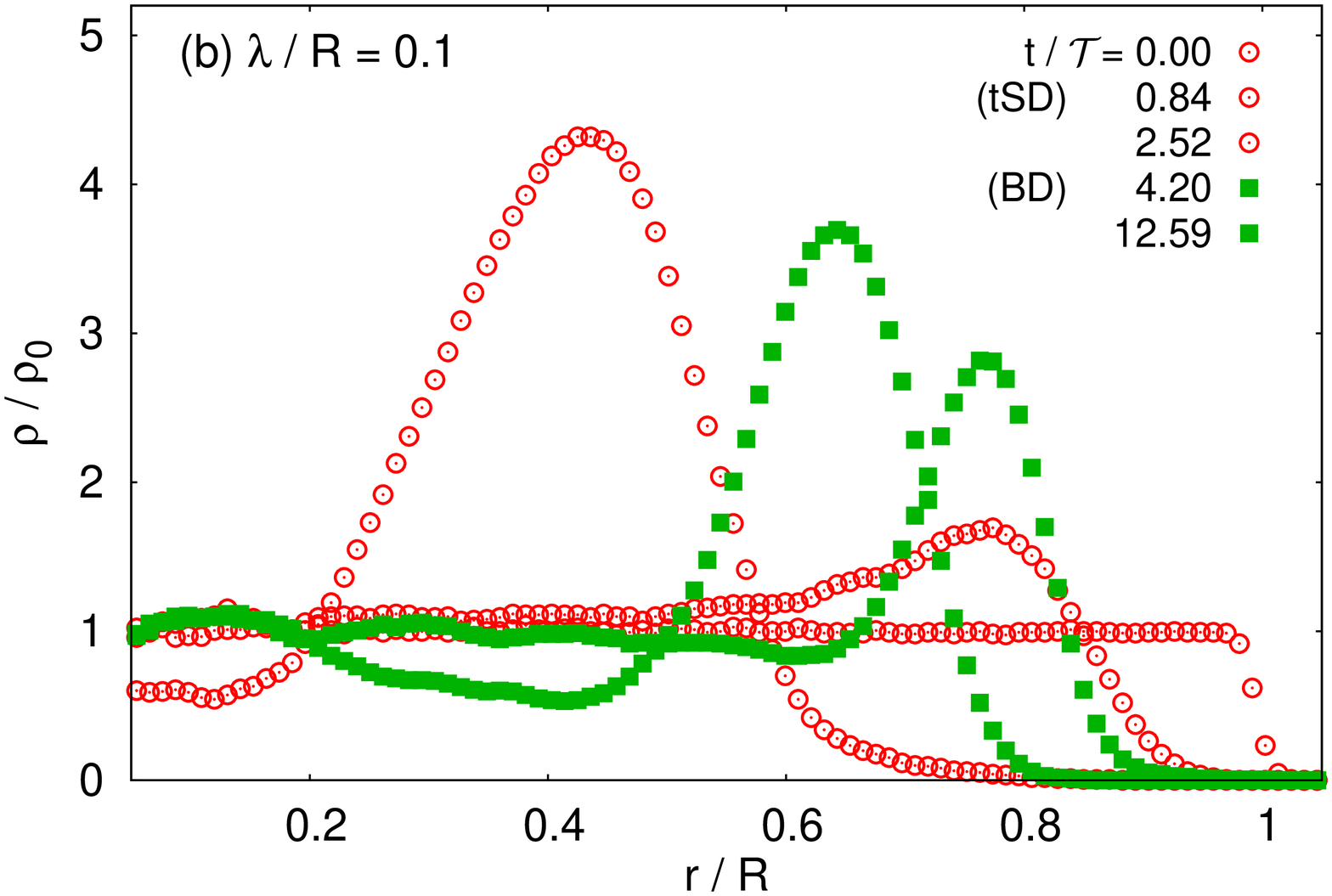,width=.7\linewidth}
  \caption{Collapse of an initial top--hat profile: (a) 2D gravity
      ($\lambda\to\infty$) without (dashed line, from theory) and with
      HI (symbols, from tSD and LB simulations of hard spheres; 2D
      close--packing density is $\varrho/\varrho_0\sim 7$). Due to
      computational costs, the LB results are derived from an average
      over just 30 runs of $50$ particles (diameter $10$ lattice units
      (l.u.)) distributed in a disk of radius $R=100$ l.u. in a box of
      size $256\times 256 \times 64$ l.u.$^3$; {the interface is
        always located {at the
          {center} of the box.}}
     {(b) Screened capillary attraction ($\lambda<\infty$) without
      (green symbols, from BD simulations) and with HI (red symbols,
      from tSD simulations). The simulations consist of $N=1804$ hard
      spheres (radius $10$ $\mu$m) distributed in a disk of radius
      $R=1832$ $\mu$m inside a box of size $8000\times 8000$
      $\mu$m$^2$. Other parameters: $\lambda = 0.1 R$,
      $1/\mathcal{K}=34$ $\mu$m, $\mathcal{T}=23835$ s.}
     }
   \label{fig:cold_coll}
\end{figure}

$(ii)$--Speedup of  {nonlinear} capillary collapse: 
 {A particularly simple limit case of the instability corresponds to
   $\mathcal{K}, \lambda\to \infty$ in \eq{eq:Dcap} (dubbed ``cold collapse''
  in 2D Newtonian gravity), allowing for an exact solution of the
  nonlinear evolution equations in the absence of HI~\cite{Bleibel:2011}:
  an initially homogeneous circular patch of
  particles (top--hat profile) remains top--hat during the evolution
  towards the simultaneous collapse of all the particles at the center
  at a time $t=\mathcal{T}$.  Fig.~\ref{fig:cold_coll}(a) addresses the effect
  of HI on this solution by means of tSD and LB simulations: due to
  the faster dynamics of the low--$k$ modes induced by HI, the
  collapse is accelerated and the top--hat profile is destroyed, with
  a  {faster increase of density at the center (clearly seen at time $t/\mathcal{T}=0.5$), 
  until close--packing effects become important (at $t/\mathcal{T}=0.8$) and
  halt the collapse}. The agreement between the results from tSD and
  LB is reassuring that the phenomenological effect by HI is captured
  already by the Oseen approximation. When the capillary attraction is screened
  ($\lambda < $ initial patch radius), the collapse develops a
  shockwave--like feature at the outer rim~\cite{Bleibel:L2011}. As
  illustrated by Fig.~\ref{fig:cold_coll}(b), the incorporation of HI
  does not alter this qualitative spatial structure but the collapse
  acceleration is very prominent.
}

\begin{figure}[th]
  \centering \epsfig{file=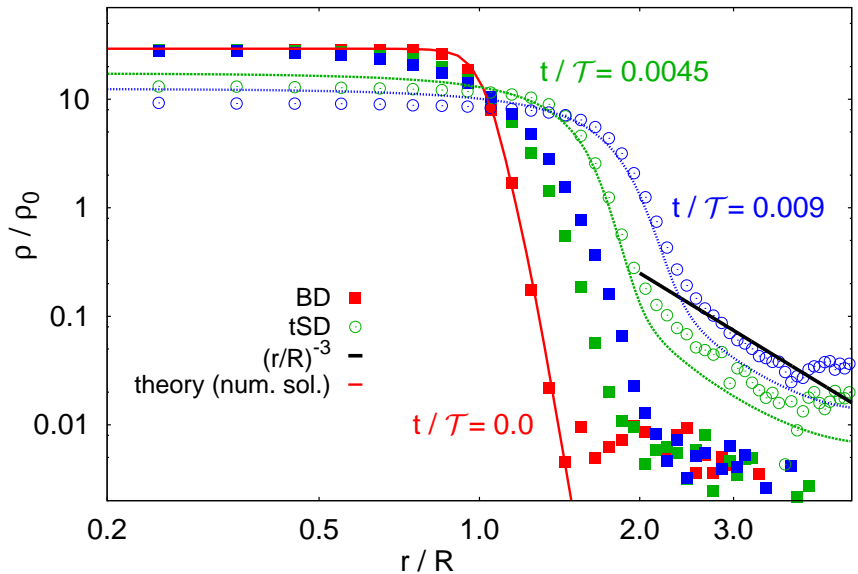,width=.7\linewidth}
  \caption{ {Diffusion of a top--hat overdensity of ideal gas
      without (filled symbols, averaged over 50000 runs of BD
      simulations) and with HI (open symbols, averaged over 10000 runs
      of tSD simulations; lines represent the numerical solution of
      Eqs.~(\ref{eq:cont})--(\ref{eq:ambient})). The simulations
      consist of $400$ particles (hydrodynamic radius $r_H=10$ $\mu$m)
      in a box of size $1000\times 1000$ $\mu$m$^2$ including an
      initial overdensity ( {$\sim 30$ $\times$ background} density) }
     { of radius $R=100$ $\mu$m. 
    } }
  \label{fig:id_gas}
\end{figure}

$(iii)$-- {Anomalous diffusion in an ideal gas: we have considered
  an initial top--hat profile
  immersed in a homogeneous background. 
  Despite the absence of (static) interactions, the
  evolution is affected by the HI if a nonvanishing hydrodynamic
  radius $r_H$ of the particles is assumed.}  This corresponds to the
idealized case of, e.g., mutually interpenetrable polymeric particles
whose radius of gyration defines $r_H$. 
 {Fig.~\ref{fig:id_gas} shows the effect of HI by means of
  simulations and the numerical solution of
  Eqs.~(\ref{eq:cont})--(\ref{eq:ambient}). One can observe how the HI
  accelerate the diffusion and induce the development of a tail
  consistent with the $r^{-3}$ decay predicted by \eq{eq:green}.}

{\em Discussion and conclusions.--}  {The singularity in $D(k)$ is
  derived from a \textit{stationary} 3D flow which affects the
  particle dynamics already at the \textit{linear} level because it is
  \textit{compressible} in the plane  {of colloidal motion}.  Thus, it is distinct from 
  the well known divergence of the diffusion coefficient
  in purely 2D systems \cite{LRW95}. 
  {The latter is} related to the Stokes paradox and due to a long--time
  tail in the 
  velocity autocorrelation, induced by the nonlinear coupling of the
  particle motion with the build-up in time of a 2D incompressible
  flow. Likewise, it differs from the divergence in lateral diffusion
  in fluid membranes,
  related also to the Stokes paradox for the 2D incompressible flow
  inside the membrane~\cite{Saffman:1975}.
  The singularity in $D(k)$ is a sole consequence of the partial
  confinement and is a quite robust result, being qualitatively
  independent of
  the specific properties (strength, range) of the effective, static
  interaction between the particles.} Thus, dynamic signatures of this
singularity 
can be expected wherever the conditions of restricted colloidal motion
and unrestricted hydrodynamic interactions are met.
 {In this regard we note that the authors of Ref.~\cite{Pesche:2000}
  also  {considered the divergence arising from  partial confinement for
  the short--time dynamics of particles confined between walls, but the association with the
  divergence in purely 2D systems 
  (by reference to Ref.~\cite{LRW95}) is misleading. 
  Possibly related to our results on {\em collective} diffusion under partial confinement
  are reports on the experimental
  observation of an enhancement of \textit{self} diffusion in
  monolayers \cite{RZMM99,Zahn:1997}  which} the authors interprete as a consequence of HI
  mediated by 3D flow.
}

 {In conclusion, under partial confinement}, i.e., colloidal motion
restricted to a plane but with hydrodynamic interactions originating
from  {3D flow} of a surrounding fluid, 
peculiar collective diffusion properties  {emerge}. Using a
mean--field model, we have identified a singularity in the long--time,
wave number dependent collective diffusion coefficient, $D(k\to 0)\sim
1/k$,
responsible for anomalous diffusion.
{The dominating dynamical effect of this singularity has been
  illustrated by simulation examples of an ideal gas (dilute limit)
  and a gas of capillary monopoles (long--ranged interparticle
  attraction)} obtained with effectively 2D Stokesian dynamics
truncated at the two--body level and with 3D
Lattice--Boltzmann{/Molecular Dynamics} simulations.

A.D. acknowledges support by the Spanish Government through Grants
No.~AIB2010DE-00263 and No.~FIS2011-24460 (partially financed by FEDER
funds). Part of the work of J.B. has been performed under the HPC-Europa2
project (project number: 228398) with the support of the European Commission -
Capacities Area - Research Infrastructure.


\end{document}